\documentclass[utf8]{FrontiersinHarvard}
\usepackage[T1]{fontenc}
\usepackage{url,hyperref,lineno,microtype,subcaption}
\usepackage[onehalfspacing]{setspace}
\usepackage{threeparttable}
\usepackage{booktabs}
\usepackage{multirow}

\def\keyFont{\fontsize{8}{11}\helveticabold }
\def\firstAuthorLast{F.\,Tonolo} %use et al only if is more than 1 author
\def\Authors{Francesca Tonolo\,$^{1,*}$}
\def\Address{$^{1}$ Molecular Physics Department, Institut de Physique de Rennes, University of Rennes - CNRS, Rennes, France}
\def\corrAuthor{Francesca Tonolo}
\def\corrEmail{Campus Beaulieu, All. Jean Perrin Bât. 11B, UMR 6251, Rennes, F-35000, France \\ francesca.tonolo.1@univ-rennes.fr \\
}

\begin{document}
\onecolumn
\firstpage{1}

\title[Collisional Excitation in Space]{Collisional Excitation in Space: Recent Advances and Future Challenges in the JWST Era} 

\author[\firstAuthorLast ]{\Authors} %This field will be automatically populated
\address{} %This field will be automatically populated
\correspondance{} %This field will be automatically populated

\extraAuth{}
\maketitle

\begin{abstract}
This perspective offers a viewpoint on how the challenges of molecular scattering investigations of astrophysical interest have evolved in recent years. Computational progress has steadily expanded collisional databases and provided essential tools for modeling non-LTE astronomical regions. However, the observational leap enabled by the JWST and new observational facilities has revealed critical gaps in these databases. In this framework, two major frontiers emerge: the characterization of collisional processes involving heavy projectiles, and the treatment of ro-vibrational excitation. The significant computational effort of these investigations emphasizes the need to test and develop robust theoretical methods and approximations, capable of extending the census of collisional coefficients required for reliable astrophysical modeling. Recent developments in these directions are outlined, with particular attention to their application and their potential to broaden the coverage of molecular systems and physical environments.

\tiny
 \keyFont{ \section{Keywords:} collision dynamics; databases; rate coefficients; non-LTE; JWST} %All article types: you may provide up to 8 keywords; at least 5 are mandatory.
\end{abstract}

\section{Introduction}
The past decade has witnessed major progress in collisional excitation studies for astrophysical applications. These advances have been decisive for interpreting astronomical observations and constraining the physical conditions of interstellar and circumstellar environments, where molecular level populations often deviate from local thermodynamic equilibrium (LTE).
In non-LTE regimes, radiative and collisional processes contribute on comparable timescales to molecular (de)-excitations. Consequently, the fundamental molecular data needed to interpret observational spectra are: energy levels, radiative rates and collisional rate coefficients. While spectroscopic measurements provide access to the first two, the determination of collisional rate coefficients largely relies on accurate quantum calculations \citep{roueff2013molecular,lique2019gas,tonolo2024ab}.

So far, state-of-the-art computational strategies, based on both an accurate treatment of the electronic motion (\emph{e.g.}, by using the gold-standard CCSD(T) method, as shown in \cite{faure2022collisional} and references therein) and the close-coupling (CC) time-independent quantum formalism \citep{arthurs1960theory} to solve the nuclear problem, have been the primary tool for producing complete datasets of collisional rate coefficients. 
%By treating both electronic and nuclear motions with advanced methods, they deliver results of high reliability for astrophysical modeling. 
For diatomic and triatomic species, typical accuracies are within 10-20\% for collisional rate coefficients \citep{lique2010rotational,yang2010communication}, and 1-10\% for pressure broadening coefficients \citep{tonolo2021improved,tonolo2025experimental,bizzocchi2024millimeter}.
In parallel, improved experimental techniques, based on molecular beams, flow-based methods and laser spectroscopy, have begun to provide very detailed information on collisions, including the observation of quantum resonances and interference effects \citep{chefdeville2013observation,brouard2014taming,bergeat2015quantum,de2020imaging,toscano2020cold,labiad2022absolute}. 
Experimental setups cannot yet probe collision dynamics across the full molecular diversity and physical conditions of astrophysical media, but they play a crucial role in testing and validating theoretical methods and approximations.

Nowadays, computed data are collected in collisional databases that are in constant growth, \emph{e.g.}, BASECOL \citep{dubernet2023basecol}, EMAA \citep{faure2025excitation} and LAMDA \citep{van2020leiden}. 
These databases now cover a large number of molecules of astrochemical interest, mostly small species (less than five atoms), including radicals and ions, interacting with the main projectiles of the interstellar medium (ISM): H$_2$, He and H. 
In most cases, rate coefficients are resolved for both fine and hyperfine structure, when present, and also include the most relevant isotopologues \citep{faure2012impact,denis2015isotopic,tonolo2022hyperfine}.
%The growing collection of collisional data has provided the %opportunity to test and validate over the years various approximations %aimed at lowering computational cost without significantly affecting %the accuracy of the outcomes. To cite a few: 
%to simplify the description of the potential, in some cases the H$_2$ %projectile can be safely assumed to be fixed in its ground rotational %state. This reduces the dimensionality of the problem, as it is %equivalent to treating H$_2$ as a structureless species, and is %particularly efficient when the target is an ionic species.
%Similarly, to ease scattering calculations when hyperfine structure is %involved, a common and very effective approximation is the recoupling %approach, first described by Alexander \& Dagdigian (1985). Widely %regarded as a reference method for treating quadrupole coupling %interactions (Faure \& Lique 2012), this approximation initially %neglects the coupling term in the molecular Hamiltonian, and %subsequently separates the spin wave-functions from the rotational %ones.

The field has made significant progress in this direction, and further developments are underway, with new datasets covering a broader range of molecules and temperature regimes. For example, collisional studies are increasingly tackling the characterization of complex organic molecules (COMs; \cite{faure2011potential,faure2019interaction,mandal2022mixed,ben2022interaction,ben2023collisional,khalifa2024collisional,ben2024rotational,dagdigian2024rotational,dagdigian2024rotationalb,demes2024first,bop2025guidelines}), defined as species with more than five atoms and at least one carbon \citep{herbst2009complex}. COMs have been detected in numerous interstellar observations, where they play a key role in astrochemical networks \citep{mcguire20222021}. Yet, their large size often makes the use of full quantum CC methods computationally prohibitive. Exact CC calculations are extremely burdensome also when addressing the hot regions of the ISM, such as hot cores or protoplanetary nebulae, where the high temperatures (often exceeding 100\,K) demand the extension of scattering calculations to a broader kinetic energy range \citep{dumouchel2010rotational,faure2016collisional,tonolo2024collisional}.
In such cases, viable strategies involve solving the nuclear Schrödinger equation with approximate quantum methods, such as coupled states (CS; \cite{mcguire1974quantum}) or infinite order sudden (IOS; \cite{pack1974space}) approximations.  
In the former one, off-diagonal Coriolis couplings are neglected in the scattering Hamiltonian, which often preserves good accuracy while reducing the computational cost by a factor of 3-10 compared to CC \citep{tobola2007rotational,lique2008quantum,troscompt2009rotational}.
The latter, based on the assumption that the time between collisions is much shorter than the rotational time, decouples different orientations of the collider. This works well at high collisional energies but may lead to physically unreliable results at low energies, where molecular rotation strongly affects cross sections. The mixed quantum/classical theory (MQCT) offers also a promising compromise in this direction, being able to extend scattering calculations to previously inaccessible systems and temperatures (\emph{e.g}, \cite{loreau2018scattering}). Developed by \cite{semenov2020mqct} and \cite{mandal2024mqct}, MQCT combines a quantum treatment of internal ro-vibrational motion with a classical description of scattering, achieving a good balance between accuracy and efficiency. At temperatures above 100\,K, MQCT results are typically within a factor of 2 with respect to CC calculations \citep{babikov2016recent,joy2024mixed}.
At high temperatures, classical treatments of the nuclear motion, as in the quasi-classical trajectory (QCT) method, are also worth to mention as valid alternatives, noticeably reducing the computational effort while providing good accuracies (\emph{e.g.}, \cite{faure2006rotational,loreau2018scattering}).

Despite major progress in methodology and the steady expansion of collisional databases, the challenges ahead continue to intensify. These are mostly driven by the rapid growth of observational data. Breakthroughs from modern astronomical facilities, such as the ALMA array of radiotelescopes \citep{wootten2009atacama}, the James Webb Space Telescope (JWST; \cite{gardner2023james}), and the SKA precursors and pathfinders \citep{rottgering2003lofar,tingay2016multi,jonas2016meerkat}, have significantly broadened the horizons of astrochemistry. 
Their unprecedented sensitivity and spectral resolution are opening new windows onto diverse environments, including planetary atmospheres, cometary comae, and exoplanets \citep{trappist2024roadmap,cordiner2024atacama,nixon2025atmosphere}. Despite steep vertical gradients in molecular densities often drive strong departures from LTE in such regions, retrieving collisional datasets to support astrochemical models remains challenging, costly, and far from straightforward.
For instance, the presence of heavy colliders such as N$_2$, CO, CO$_2$ and H$_2$O requires the development and application of advanced computational strategies. 
At the same time, infrared observations of vibrational molecular signatures, boosted by the MIRI \citep{wright2004jwst} and NIRSPEC \citep{bagnasco2007overview} instruments on JWST, are flourishing, providing unprecedented insights into the underlying chemistry. Accurately modeling collisional processes under these conditions remains a pressing frontier, driving ongoing innovation in theoretical chemistry and molecular physics.

In the following, I will focus on these emerging frontiers. From my perspective, these research areas will play a crucial role in keeping pace the rapid observational advances brought by the new generation of astronomical instruments and by \emph{in situ} space missions. In~\S\,\ref{section2}, current investigations of molecular de-excitation induced by collisions with heavy partners (\emph{i.e.}, species heavier than H$_2$) are presented, while ~\S\,\ref{section3} discusses the methodologies of particular relevance for treating ro-vibrational excitation. ~\S\,~\ref{section4} concludes with a summary and outlook for future developments.

\section{Heavy Colliders}
\label{section2}
Most of collisional investigations of astrophysical interest address (de-)excitation processes induced by H$_2$ and He, the most abundant projectiles in the ISM. Yet, recent observational advances are revealing environments with far greater chemical diversity. 

A representative example is cometary comae, the gaseous envelopes that sublimate from cometary ices under solar radiation \citep{bockelee2008cometary,roth2021leveraging,biver2022observations,cordiner2023gas}. The composition of cometary comae carries the chemical legacy of the early stages of Solar System formation, preserved in cometary ices and released through sublimation, and provides clues on the potential delivery of prebiotic molecules to Earth \citep{matthews2008hydrogen,mumma2011chemical,jung2013mechanisms}.
However, the low densities of cometary comae (1\,$<$\,\emph{n}\,$<$\,10$^{10}$\,cm$^{-3}$) place strong constraints on astrochemical modeling, unless supported by datasets of collisional coefficients with the dominant perturbers \citep{loreau2022effect}. The identity of these projectiles depends on the heliocentric distance of the comet, as solar energy selectively drives molecular sublimation. CO dominates at large heliocentric distances, evolving into CO\,/\,CO$_2$ mixtures at intermediate ranges. At $\sim$1\,au, the solar radiation is intense enough to sublimate H$_2$O from the nucleus, making it the dominant gas. Here, collisions with free electrons have also an impact, which particularly intensifies in the intermediate regions of the coma, at cometocentric distances between $10^2$ and $10^4$\,km \citep{loreau2022effect}.

Titan's atmosphere provides another illustrative case. 
Titan is the only other body in the Solar System, besides Earth, with a dense, N$_2$-rich atmosphere and features reminiscent of primordial Earth. 
Here, strong vertical abundance gradients, especially above 800 km in the thermosphere, produce pronounced non-LTE effects \citep{rezac2013rotational,nixon2025atmosphere}. 
Similar conditions are found in the water vapor dominated atmospheres of Jupiter's Moons Ganymede, Calisto, and Europa, where non-LTE effects peak at sub-solar points due to intense sublimation and photo-desorption processes \citep{mogan2021tenuous, enya2022ganymede, vorburger20223d}.
To conclude, non-LTE effects are also well documented in the CO$_2$ rich atmospheres of Mars \citep{johnson1976nonthermal,piccialli2016co2} and Venus \citep{lopez2007non,gilli2009limb}.

These examples underscore the critical role of collisional rate coefficients in modeling chemically diverse environments. The presence of heavy colliders, however, introduces substantial challenges for scattering calculations, both methodological and computational. Indeed, the dense rotational structure of N$_2$, CO, CO$_2$ and H$_2$O multiplies the number of collisional channels to be considered, making full quantum scattering calculations unfeasible in most cases. 
\begin{table}[h!]
    \centering
    \renewcommand{\arraystretch}{1.0}
    \setlength{\tabcolsep}{5pt} 
    \caption{State-of-the-art datasets of collisional rate coefficients relevant for modeling cometary comae, with corresponding references.}
    \vspace{2mm}
    \label{tab:heavycoll}
    \small
    \begin{tabular}{c c c c c c}
        \toprule
          \multicolumn{1}{c}{System\textsuperscript{a}} & \multicolumn{1}{c}{State-of-the-art} & \multicolumn{1}{c}{Method} & \multicolumn{1}{c}{Temperature} & \multicolumn{1}{c}{Dataset} 
          %&\multicolumn{1}{c}{Rotational basis} 
          & \multicolumn{1}{c}{Other data\textsuperscript{b}}\\
        \midrule
\multirow{2}{*} {CO $-$ CO} & \multirow{2}{*} {\cite{zoltowski2023excitation}} & \multirow{2}{*} {CS} & \multirow{2}{*} {5-150\,K} & State-to-state  %& \multirow{2}{*} {$J(\text{CO})=10$}  
& \multirow{2}{*} {Refs. 1$-$7} \\
 &  &  &  & Thermalized\textsuperscript{c}    &  \\
\multicolumn{6}{c}{~} \\

\multirow{2}{*} {CO $-$ H$_2$O} & \multirow{2}{*} {\cite{faure2020effect}} & \multirow{2}{*} {SACM} & \multirow{2}{*} {5-100\,K} & State-to-state  %&  $J(\text{CO})=10$  
& \multirow{2}{*} {Refs. 8$-$9} \\
 &  &  &  & Thermalized\textsuperscript{c}  %&$J(\text{H$_2$O})=7$ 
 &  \\
\multicolumn{6}{c}{~} \\

\multirow{2}{*} {HCN $-$ CO} & \multirow{2}{*} {\cite{tonolo2025collisional}} & \multirow{2}{*} {SACM} & \multirow{2}{*} {5-50\,K} & State-to-state  %&  $J(\text{HCN})=9$  
& \multirow{2}{*} {~} \\
 &  &  &  & Thermalized\textsuperscript{c}  %& $J(\text{CO})=8$ 
 &  \\
\multicolumn{6}{c}{~} \\

\multirow{3}{*} {HCN $-$ H$_2$O} & \multirow{3}{*} {\cite{zoltowski2025collisional}} & \multirow{3}{*} {SACM} & \multirow{3}{*} {10-100\,K} & State-to-state  %&  $J(\text{HCN})=10$  
& \multirow{3}{*} {Ref. 10} \\
 &  &  &  & Thermalized\textsuperscript{c}  %& $J(\text{H$_2$O})=2$ 
 &  \\
 &  &  &  & Ground\textsuperscript{d}   &  \\
\multicolumn{6}{c}{~} \\

\multirow{2}{*} {CS $-$ CO} & \multirow{2}{*} {\cite{godard2025promising}} & \multirow{2}{*} {SACM} & \multirow{2}{*} {5-30\,K} & State-to-state  %&  $J(\text{CS})=9$  
& \multirow{2}{*} {~} \\
 &  &  &  & Thermalized\textsuperscript{c}  %& $J(\text{CO})=5$ 
 &  \\
\multicolumn{6}{c}{~} \\

\multirow{2}{*} {CS $-$ H$_2$O} & \multirow{2}{*} {Godard et al., \emph{in prep.}} & \multirow{2}{*} {SACM} & \multirow{2}{*} {5-100\,K} & State-to-state  %&  $J(\text{CS})=18$  
& \multirow{2}{*} {~} \\
 &  &  &  & Thermalized\textsuperscript{c}  %& $J(\text{H$_2$O})=3$ 
 &  \\
\multicolumn{6}{c}{~} \\

\multirow{2}{*} {HF $-$ H$_2$O} & \multirow{2}{*} {\cite{loreau2022effect}} & \multirow{2}{*} {SACM} & \multirow{2}{*} {5-150\,K} & State-to-state  %&  $J(\text{HF})=6$  
& \multirow{2}{*} {~} \\
 &  &  &  & Thermalized\textsuperscript{c}  %& $J(\text{H$_2$O})=6$ 
 &  \\
\multicolumn{6}{c}{~} \\

\multirow{2}{*} {H$_2$O $-$ H$_2$O} & \multirow{2}{*} {\cite{mandal2023improved}} & \multirow{2}{*} {MQCT} & \multirow{2}{*} {5-1000\,K} & State-to-state  %& \multirow{2}{*} {$J(\text{H$_2$O})=7$}  
& \multirow{2}{*} {Refs. 11$-$14} \\
 &  &  &  & Thermalized\textsuperscript{c}    &  \\
\multicolumn{6}{c}{~} \\

    \bottomrule
        
    \end{tabular}
    \vspace{2mm}
    \\
    \raggedright{\textsuperscript{a} Target species $-$ collider.} 
    \\
    \raggedright{\textsuperscript{b} If present, previously computed dataset are referenced as: (1) \cite{bostan2024mixed}; (2) \cite{ndengue2015rotational} ; (3) \cite{phipps2002investigation}; (4) \cite{sun2020molecular}; (5) \cite{laskowski2022rotational}; (6) \cite{cordiner2022sublime}; (7) \cite{zoltowski2022collisional}; (8) \cite{green1993collisional}; (9) \cite{biver1999spectroscopic}; (10) \cite{dubernet2019first}; (11) \cite{mandal2023rate}; (12) \cite{boursier2020new}; (13) \cite{semenov2017mqct}; (14) \cite{buffa2000h2o}.}
    \\
    \raggedright{\textsuperscript{c} Thermalized rate coefficients over the rotational population of the collider at each kinetic temperature.} 
    \\
    \raggedright{\textsuperscript{d} The collider is kept fixed at its ground rotational state.}    
\end{table}

%In recent years, alternative approaches has been developed that achieve reliable accuracy while drastically reducing computational cost.
Recently, the Statistical Adiabatic Channel Model (SACM), originally introduced by \cite{quack1975complex} and later refined by \cite{loreau2018efficient}, showed particular promise for very low-temperature environments such as cometary atmospheres.
While drastically reducing the computational cost, the SACM is particularly accurate for systems that form a stable intermediate complex, allowing for a statistical energy redistribution.
%The SACM neglects the nuclear kinetic contribution from the interaction Hamiltonian, drastically reducing the computational cost.
%In terms of accuracy, the SACM is particularly effective for systems that form a stable intermediate complex, allowing for a %statistical energy redistribution.
This makes it especially effective for collisions among molecules forming stable complexes, such as in ionic interactions or collisions between strongly polar species \citep{loreau2018efficient,loreau2018scattering,balancca2020inelastic,pirlot2025collisional}. For cometary applications, the SACM has already been tested on several systems, typically deviating by less than a factor of 2 from exact CC results for rate coefficients thermalized with respect to the rotational temperature of the projectile \citep{godard2025promising, tonolo2025collisional}.
For weakly bound systems such as CO$-$CO, however, statistical methods loose their hold. These systems involve instead a relatively small number of asymptotically closed channels, making them suitable for more computationally demanding methods, like the quantum CS approach \citep{zoltowski2022collisional}.
While SACM and CS remain the most practical tools for the low temperature regions ($\sim$5-100\,K), their applicability rapidly decreases at higher temperatures, as in planetary atmospheres or comets near the Sun. At elevated temperatures, indeed, the lifetime of the collisional complex is too short for SACM to remain valid, whereas the large number of channels required by CS makes the calculations computationally prohibitive.
Under these conditions, MQCT offers a valid alternative \citep{mandal2023rate,mandal2023improved}. 
\begin{figure}[h!]
\begin{center}  \includegraphics[scale=0.53]{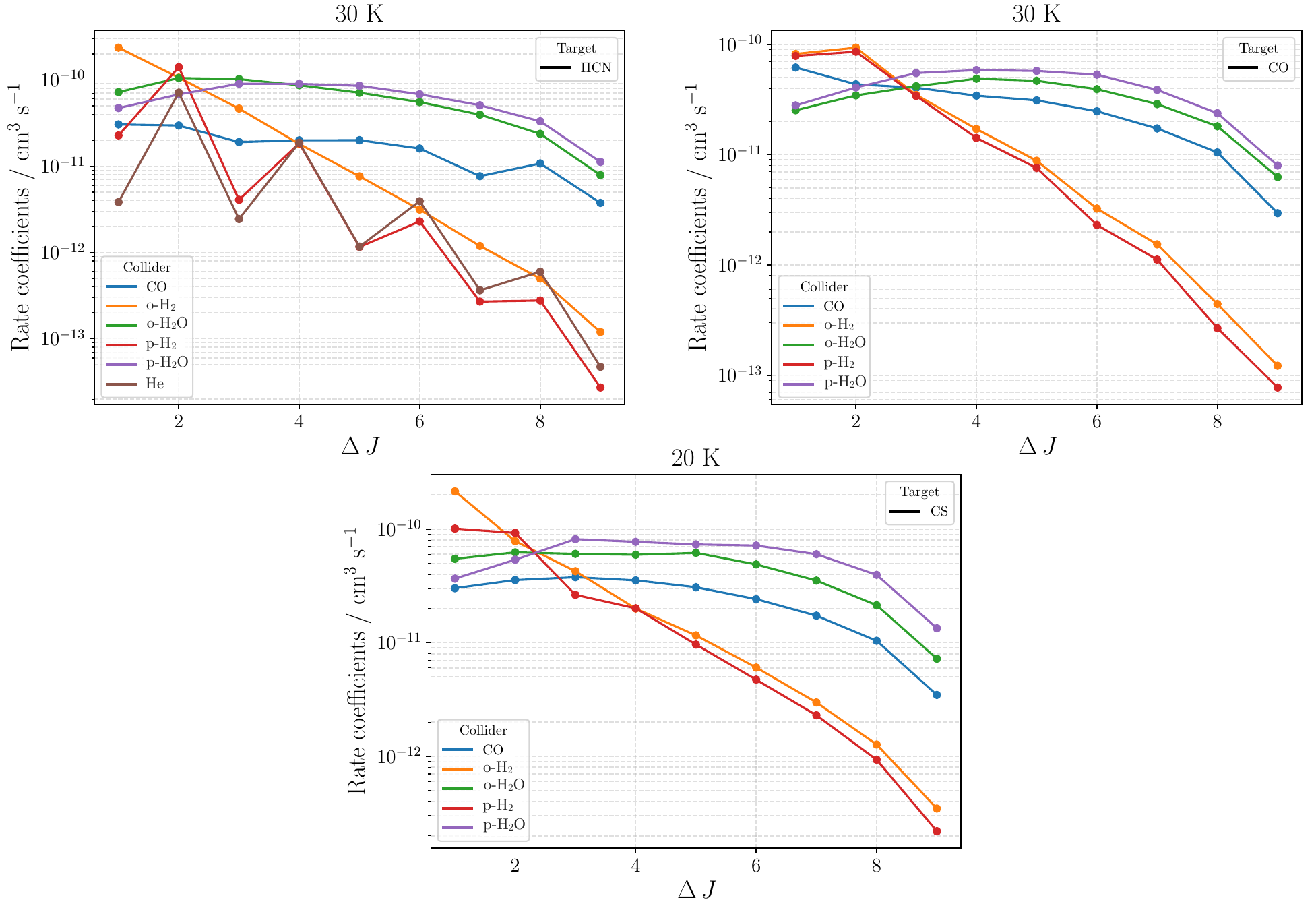}
\end{center}
\caption{Comparison of a reduced set of rate coefficients (cm$^3$\,s$^{-1}$) for the de-excitation starting from the $J=9$ rotational level of HCN, CO and CS by different projectiles, as a function of $\Delta J$.}
\label{fig1}
\end{figure}

Despite recent progress, collisional datasets involving heavy colliders remain scarce. Table~\ref{tab:heavycoll} summarizes the few available cases of astrophysical interest involving pure rotational transitions, together with the methods employed and their ranges of applicability. 

Faced with this paucity of data, a common workaround is to use existing datasets for the same molecular target to approximate its behavior with missing colliders. To test this assumption, Figure~\ref{fig1} compares a reduced set of collisional rate coefficients for the rotational de-excitation of CO \citep{faure2020effect,zoltowski2022collisional,dagdigian2022collisional}, HCN \citep{dumouchel2010rotational,hernandez2017rotational,tonolo2025collisional,zoltowski2025collisional}, and CS (\cite{denis2018new,godard2025promising}; Godard et al., in prep.), starting from $J=9$, by different projectiles. 
Two key trends emerge. First, heavy colliders such as CO and H$_2$O drive dynamics that differ significantly from those induced by typical ISM projectiles, confirming the poor reliability of the latter as templates for cometary gases.
Second, the effect of cometary projectiles appears strongly system-dependent: for weakly polar targets like CO, rate coefficients with H$_2$O and CO agree within a factor of 3, whereas for more polar molecules the deviations grow, reaching up to a factor of 8 for HCN (2.97\,D). 
This underlines a key limitation: CO is an unreliable proxy for H$_2$O, and \emph{viceversa}, especially for polarized molecules whose collisional response is highly sensitive to the nature of the collider.
%Second, for all the three targets, no substantial differences are found between collisions with the \emph{ortho} and \emph{para} states of H$_2$O. This contrasts with the case of HCN colliding with H$_2$, where a strong decrease in propensity was observed for odd $\Delta J$ transitions induced by \emph{ortho}-H$_2$. The same propensity rules emerges when He is used as a collider, consistent with its same spherical symmetry. 
These findings remain preliminary, given the limited datasets and heterogeneous accuracy of available methods. A systematic benchmarking effort will be essential to establish robust reference datasets.

\section{Ro-vibrational Excitation}
\label{section3}
The advent of JWST, together with the high resolution of new generation of observational facilities, have lifted the veil on ro-vibrational spectroscopy as a diagnostic tool of scarcely explored astrophysical environments, ranging from warm shocked gas and interstellar hot cores to planetary and exoplanetary atmospheres \citep{ahrer2023jwst, espinoza2025highlights, van2025jwst}. In these regions, vibrationally excited states are significantly populated.
Ro-vibrational spectra also reveal information inaccessible to purely rotational diagnostics: molecules without permanent dipole moments, such as CH$_4$, C$_2$H$_2$, and CO$_2$, become observable when vibrational excitation breaks their intrinsic symmetry \citep{lacy2013interpretation}. 
%In addition, infrared transitions among vibrational levels can significantly enhance the intensity of rotational lines through infrared pumping \citep{carroll1981infrared}, boosting their detectability. 
%An observational turning point in this direction has been marked by JWST for %exoplanetary science \citep{ahrer2023jwst}, which, since its launch in 2021, has %already collected hundreds of infrared observations very high signal-to-noise, %resolution, and broad wavelength coverage. These observations are significantly %advancing our understanding of exoplanetary atmospheres (Espinoza and Perrin, %2025, and references therein), and are attracting increasing interest from the %scientific community.
However, non-LTE effects are particularly pronounced under these conditions \citep{kristensen2012water,fossati2021non, borsa2022high, wright2022non, bowesman2025tiramisu}, with deviations from thermal equilibrium often exceeding those seen in purely rotational populations (see \cite{bruderer2015ro,van2024joys+} for illustrative cases of HCN and SO$_2$). 
Moreover, the large radiative de-excitation rates of ro-vibrational transitions substantially raise the critical densities required for vibrational thermalization. Consequently, assuming vibrational LTE at the gas kinetic temperature may lead to misinterpretations. Accurate datasets of collisional rate coefficients are therefore essential for reliable radiative transfer models of these environments.

In the past, vibrational and ro-vibrational collisional investigations were conducted mainly relying on QCT \citep{aoiz1992quasiclassical,varandas1994method}, and quantum-classical (QC, or semiclassical; \cite{billing1984semiclassical,billing1987rate}) methods. While the former one uniquely relies on classical dynamics treatments, the latter lays the same theoretical foundations of MQCT, and describes molecular rotations and translational motion in a classical way, while introducing a quantum description of the vibrational degrees of freedom. QC methods have undergone significant refinements in recent years
\citep{hong2023inelastic,yang2025quantum} and are now widely employed to expand collisional databases, typically covering very high temperatures (up to $\sim$9000\,K). 
Large databases of rate coefficients of pure vibrational transitions are currently available, often involving heavy colliders such as N$_2$ and CO (\emph{e.g.}, \cite{billing2003sensitivity,kurnosov2010database,lombardi2019full,hong2020inelastic,hong2023improved}). 
Collisional databases accounting for the rotational structure among each vibrational state, on the other hand, remain comparatively scarce, though they are in a steady expansion (the reader is referred to \cite{hong2021energy,hong2021vibrational,yang2025quantum}, and references therein). 

However, the lack of a proper quantum treatment remains a major limitation, preventing high accuracy below 200\,K and thereby reducing the reliability of such data for most astrophysical applications.
On the other hand, quantum investigations that fully account for ro-vibrational degrees of freedom have so far been limited to very small systems, typically involving only diatomic molecules (\emph{e.g.}, \cite{alexander1973fully, schaefer1975theoretical, ramaswamy1977vibration, balakrishnan2000vibrational, tao2007role, klos2008theoretical, stewart2010rovibrational}). 
The vibrational modes that typically involve lower frequencies, and are therefore more easily populated (leading to stronger observational lines) are bending modes or large-amplitude modes such as the CH$_3$ internal rotation (torsion mode).
However, only a handful of full quantum datasets exists for molecules with more than two atoms, as summarized in Table~\ref{tab:vib}. 
This limitation reflects the methodological and numerical challenges of computing collision-induced ro-vibrational excitation. Such calculations require higher dimensional treatment of the potential, large vibrational basis sets, and the evaluation of complex multidimensional integrals to obtain the scattering matrix. 
%Another critical gap concerns pure rotational transitions in vibrationally excited states, which are generally assumed to be identical to the ground-state values. However, this assumption has yet to be validated, despite evidence that other collisional properties, such as pressure broadening (CITA ALTRI) \citep{tonolo2025experimental}, vary significantly with vibrational excitation.
\begin{table}[h!]
    \centering
    \renewcommand{\arraystretch}{1.0}
    \setlength{\tabcolsep}{8pt} 
    \caption{State-of-the-art datasets of collisional rate coefficients for polyatomic molecules, which include a quantum treatment of ro-vibrational excitation for selected vibrational modes, with corresponding references.}
    \vspace{2mm}
    \label{tab:vib}
    \small
    \begin{tabular}{c c c c c}
        \toprule
          \multicolumn{1}{c}{System\textsuperscript{a}} & \multicolumn{1}{c}{State-of-the-art} & \multicolumn{1}{c}{Method} & \multicolumn{1}{c}{Vibrational modes} & \multicolumn{1}{c}{Other data\textsuperscript{b}}\\
        \midrule
\multirow{1}{*}{CH$_2$ $-$ He} & \multirow{1}{*}{\cite{ma2014theoretical}} & \multirow{1}{*}{CC} & bending & \multirow{1}{*}{Refs. 1$-$2} \\
\multicolumn{5}{c}{~} \\

\multirow{1}{*}{CH$_3$ $-$ He} & \multirow{1}{*}{\cite{ma2013theoretical}} & \multirow{1}{*}{CC} & umbrella mode & \multirow{1}{*}{Refs. 3$-$4} \\
\multicolumn{5}{c}{~} \\

\multirow{1}{*}{HCN $-$ He} & \multirow{1}{*}{\cite{stoecklin2013ro}} & \multirow{1}{*}{CC} & bending & \multirow{1}{*}{Ref. 5} \\
\multicolumn{5}{c}{~} \\

\multirow{1}{*}{DCN $-$ He} & \multirow{1}{*}{\cite{denis2015isotopic}} & \multirow{1}{*}{CC} & bending & \multirow{1}{*}{~} \\
\multicolumn{5}{c}{~} \\

\multirow{1}{*}{C$_3$ $-$ He} & \multirow{1}{*}{\cite{stoecklin2015rovibrational}} & \multirow{1}{*}{CC} & bending & \multirow{1}{*}{Ref. 6} \\
 \multicolumn{5}{c}{~} \\

\multirow{1}{*}{H$_2$O $-$ H$_2$} & \multirow{1}{*}{\cite{garcia2024bending}} & \multirow{1}{*}{CC} & bending & \multirow{1}{*}{Refs. 7-11} \\
 \multicolumn{5}{c}{~} \\

\multirow{1}{*}{H$_2$O $-$ He} & \multirow{1}{*}{\cite{stoecklin2021close}} & \multirow{1}{*}{CC} & bending & \multirow{1}{*}{~} \\
 \multicolumn{5}{c}{~} \\

%\multirow{2}{*}{H$_2$O $-$ e$^-$} & \multirow{2}{*}{\cite{ayouz2024theoretical}} %& \multirow{2}{*}{\emph{ab initio}} & stretching & \multirow{2}{*}{Refs. 12-13} \\
% & &  & bending & \\
% \multicolumn{5}{c}{~} \\

\multirow{1}{*}{NH$_3$ $-$ He} & \multirow{1}{*}{\cite{loreau2024vibrational}} & \multirow{1}{*}{CC} & umbrella mode & \multirow{1}{*}{Ref. 12} \\
 \multicolumn{5}{c}{~} \\

\multirow{2}{*}{CO$_2$ $-$ He} & \multirow{1}{*}{\cite{selim2021multi}} & \multirow{1}{*}{CC/MC-DWBA} & stretching & \multirow{2}{*}{Ref. 13} \\
 & \multirow{1}{*}{\cite{selim2023state}} & \multirow{1}{*}{CC} & bending & \\
 \multicolumn{5}{c}{~} \\

\multirow{1}{*}{CH$_3$OH $-$ He} & \multirow{1}{*}{\cite{rabli2011rotationally}} & \multirow{1}{*}{CC} & torsion & Ref. 14 \\
 \multicolumn{5}{c}{~} \\
 
 \bottomrule
        
\end{tabular}
    \vspace{2mm}
    \\
    \raggedright{\textsuperscript{a} Target species $-$ collider.} 
    \\
    \raggedright{\textsuperscript{b} If present, previously computed dataset are referenced as: (1) \cite{ma2012theoretical}; (2) \cite{ma2011theoretical}; (3) \cite{dagdigian2011theoretical}; (4) \cite{dagdigian2013theoretical}; (5) \cite{denis2013interaction}; (6) \cite{denis2014rovibrational}; (7) \cite{faure2005role}; (8) \cite{faure2005full}; (9) \cite{valiron2008r12}; (10) 
    \cite{stoecklin2019rigid} ; (11)
    \cite{wiesenfeld2022quenching}; (12) \cite{loreau2015scattering}; (13) \cite{selim2022efficient}; (14)
    \cite{pottage2003torsional}.}  
\end{table}

An efficient way out in this direction was proposed by \cite{selim2021multi}, who treated all rotational channels for each vibrational quantum number with the full CC method, while handling the weaker couplings between vibrational states perturbatively with the multi-channel distorted-wave Born approximation (MC-DWBA). This strategy reduces CPU time by a factor of 3 without compromising accuracy. Nevertheless, such gain remains insufficient to address larger systems and broader energy ranges.
To further reduce the computational effort, the same authors benchmarked the CS approximation against CC results, with also two improved variants that include Coriolis couplings to first order, \emph{i.e.}, the nearest-neighbor Coriolis coupling approximation (NNCC; \cite{yang2018improved,selim2022efficient}). For the CO$_2$–He system, NNCC surpassed standard CS in reproducing CC results.
In terms of CPU and memory requirements, NNCC offered best balance between efficiency and accuracy for both inelastic cross sections and rate coefficients. Moreover, combining MC-DWBA with NNCC yielded similar improvements in computational efficiency as with CC, without loss of accuracy.

More recently, the same authors tested the VCC-IOS approximation for state-to-state ro-vibrational transitions of CO$_2$ colliding with He \citep{selim2023state}. Originally developed by Clary and co-workers in the 1980s \citep{clary1981quantum, clary1982ab, clary1995mechanisms, banks1987coupled, wickham1987experimental}, VCC-IOS combines a CC treatment of vibrational motion with the IOS approximation for rotations, drastically reducing the computational cost. This method has long proven effective for vibrational quenching and ro-vibrational (de)-excitation in diatomic systems \citep{lique2006ro, lique2007ro, tobola2008ro, balancca2017ro}, but its applicability to polyatomic molecules was partially explored. The results showed a strong dependence on vibrational frequency: for high frequency modes (symmetric and asymmetric CO$_2$ stretching), VCC-IOS deviated from CC by up to three orders of magnitude \citep{selim2021multi}. In contrast, for the low-frequency bending mode of CO$_2$, the method reproduced collisional coefficients within $\sim$50\% of CC values \citep{selim2023state}. Although this behavior may seem counterintuitive, given the strong coupling between rotational and bending motions, it can be rationalized by the improved performance of the VCC-IOS approach as the magnitude of the cross sections increases. Benchmark calculations assessing this dependence in other collisional systems would provide valuable insights.

When addressing bending modes of triatomic molecules, another promising strategy was proposed by \citet{denis2013interaction}: the rigid-bender averaged approximation (RBAA). Here, the interaction potential is averaged over the bending wave function before scattering calculations, effectively reducing the problem to an atom colliding with a linear molecule whose rotation and bending motions are decoupled.
Although drastic, this approximation shows promise for extending ro-vibrational collisional studies to medium- and large-sized polyatomic systems \citep{stoecklin2013ro}.

Finally, it is worth mentioning the significant progress achieved over the years in predicting the vibrational excitation of polyatomic molecules by electron impact. A paradigmatic case addresses the H$_2$O molecule, which has long served as a benchmark system for testing and validating various theoretical approaches \citep{nishimura1995electron,curik2003vibrationally,nishimura2004vibrational,song2021cross}. Building upon these studies, \cite{faure2008collisional} derived for the first time ro-vibrational rate coefficients under the assumption of a complete decoupling between rotational and vibrational motions. More recently, \cite{ayouz2021cross,ayouz2024theoretical} have developed and refined an \emph{ab initio} methodology capable of reproducing vibrational thermalized rate coefficients with remarkable accuracy over a broad temperature range (10–10000\,K), encompassing the 13 lowest energy levels of H$_2$O. The extension of these datasets to rotationally resolved vibrational rate coefficients is currently in progress.

\section{Discussion}
\label{section4}
The recent advances in computational methods have considerably expanded collisional databases, enabling progress on several long-standing challenges. Yet, this is only the tip of the iceberg. 
%Much remains to be done, especially since the advent of JWST and other high-resolution facilities, which profoundly reshaped the observational landscape. 
The breakthroughs delivered by the JWST have raised the stakes, providing new observational data from complex environments such as cometary comae, planetary and exoplanetary atmospheres, and warm shocked regions. In all these regions, strong departures from LTE are frequently observed, highlighting the need for accurate collisional rate coefficients to support radiative transfer models. 

In this perspective, two major frontiers have been identified and discussed. 
The first concerns collisions involving heavy colliders. Here, the SACM approach appears promising to address systems forming stable intermediates at low temperatures. The CS approximation is more suitable for weakly bound systems, while MQCT methods provide an effective strategy to extend calculations to higher temperatures. 
The second frontier involves the treatment of ro-vibrational excitation. Hybrid strategies, such as combining CC with perturbative schemes or employing improved variants of the CS approximation, have shown encouraging results.
Together, these developments outline a toolkit of methods that, if carefully benchmarked, can help address these challenges.

Still, there is room for improvement. The systems studied so far are few, and methodological benchmarks remain fragmented or incomplete. Even when new methods show promise on a given target, their general applicability remains untested. 

Expanding collisional databases is therefore an urgent priority. This will require coordinated efforts on multiple fronts: developing and testing approximate quantum approaches, benchmarking them systematically against full quantum results, and producing datasets covering the relevant physical conditions of planetary, cometary, and exoplanetary environments. Machine learning may also play a pivotal role in this regard, offering efficient ways to interpolate and extrapolate collisional data, reduce computational cost, and uncover transferable patterns across molecular systems (the reader is referred to \cite{mihalik2025accurate} and reference therein for some illustrative examples). Data-driven models trained on accurate quantum calculations could accelerate the expansion of collisional databases, extending them to larger systems and broader physical regimes.

This perspective is not intended as an exhaustive review of past work. Rather, its goal is to emphasize the evolving challenges in the field and to suggest possible directions for addressing them in the coming years.

\section*{Conflict of Interest Statement}
%All financial, commercial or other relationships that might be perceived by the academic community as representing a potential conflict of interest must be disclosed. If no such relationship exists, authors will be asked to confirm the following statement: 
The authors declare that the research was conducted in the absence of any commercial or financial relationships that could be construed as a potential conflict of interest.

\section*{Author Contributions}
The author confirms being the sole contributor of this work and has approved it for publication.
%The Author Contributions section is mandatory for all articles, including articles by sole authors. If an appropriate statement is not provided on submission, a standard one will be inserted during the production process. The Author Contributions statement must describe the contributions of individual authors referred to by their initials and, in doing so, all authors agree to be accountable for the content of the work. Please see  \href{https://www.frontiersin.org/about/policies-and-publication-ethics#AuthorshipAuthorResponsibilities}{here} for full authorship criteria.

\section*{Funding}
This work received the support by the “Programme National de Planétologie” (PNP) under the responsibility of INSU, CNRS (France).

\section*{Acknowledgments}
I would like to acknowledge Prof. François Lique and Dr. Paul Pirlot Jankowiak for the fruitful discussions and the careful reading and insightful comments for this manuscript.

%\section*{Supplemental Data}
% \href{http://home.frontiersin.org/about/author-guidelines#SupplementaryMaterial}{Supplementary Material} should be uploaded separately on submission, if there are Supplementary Figures, please include the caption in the same file as the figure. LaTeX Supplementary Material templates can be found in the Frontiers LaTeX folder.

\section*{Data Availability Statement}
The datasets analyzed in this study can be found in the EMAA database \citep{faure2025excitation}, \url{https://emaa.osug.fr}. Further inquiries can be directed to the corresponding author.

% Please see the availability of data guidelines for more information, at https://www.frontiersin.org/about/author-guidelines#AvailabilityofData

\bibliographystyle{Frontiers-Harvard} 
\bibliography{test}

%%% Make sure to upload the bib file along with the tex file and PDF
%%% Please see the test.bib file for some examples of references

\end{document}